\documentclass[universe,preprints,oneauthor,pdftex,10pt,a4paper]{mdpi} 

\firstpage{1} 
\articlenumber{x}
\doinum{10.3390/------}
\pubvolume{xx}
\pubyear{2016}
\copyrightyear{2016}
\externaleditor{Academic Editor: name}
\history{Received: date; Accepted: date; Published: date}

\pdfoutput=1

\Title{Non-standard hierarchies of the runnings of the spectral index in inflation}

\Author{Chris Longden $^{1}$}
\AuthorNames{Chris Longden}

\address{%
$^{1}$ \quad University of Sheffield, United Kingdom; cjlongden1@sheffield.ac.uk}

\abstract{Recent analyses of cosmic microwave background surveys have revealed hints that there may be a non-trivial running of the running of the spectral index. If future experiments were to confirm these hints, it would prove a powerful discriminator of inflationary models, ruling out simple single field models. We discuss how isocurvature perturbations in multi-field models can be invoked to generate large runnings in a non-standard hierarchy, and find that a minimal model capable of practically realising this would be a two-field model with a non-canonical kinetic structure. We also consider alternative scenarios such as variable speed of light models and canonical quantum gravity effects and their implications for runnings of the spectral index.}


\conferencetitle{VARCOSMOFUN'16}

\usepackage{amsmath}
\usepackage{amssymb} 
\usepackage{graphicx} 
\usepackage[utf8]{inputenc}
\usepackage{siunitx}

\newcommand{\beq}{\begin{equation}}
\newcommand{\eeq}{\end{equation}\\}
\newcommand{\rpar}[1]{\left(#1\right)}
\newcommand{\spar}[1]{\left[#1\right]}
\newcommand{\bd}{{\rm d}}

\newcommand{\TRS}{\mathcal{T}_\mathcal{RS}}
\newcommand{\TSS}{\mathcal{T}_\mathcal{SS}}

\DeclareSIUnit\parsec{pc}

\begin{document}



\section{Introduction} \label{sec:intro}

Cosmic inflation, a period of accelerated expansion in the very early universe, was originally proposed as a solution to the flatness, horizon and monopole problems of the standard hot big bang cosmology \cite{Guth:1980zm,Linde:1981mu}, but its main success today is arguably its explanation of the primordial density fluctuations which served as the seeds of structure formation in our universe. Inflation's generic prediction of a nearly scale invariant spectrum of primordial fluctuations has established it as a widely accepted element of early universe cosmology, though alternatives such as variable speed of light scenarios have been proposed. Despite the apparent need for an accelerating expansion in the early universe, the underlying mechanism responsible for causing it remains unclear. Exotic matter such as scalar fields, possibly motivated by supersymmetric or string theories, and modifications of General Relativity are amongst the vast number of models \cite{Escudero:2015wba,Martin:2013tda} seeking to explain inflation.

Experimental measurements of the anisotropies in the cosmic microwave background (CMB) radiation can be used to constrain the details of the spectrum of primordial fluctuations, the predictions for which differ from model to model. This can, in principle, let us narrow down the range of feasible models, and indeed, recent advances in this direction, particularly with the Planck and BICEP missions have begun to rule out the most simplistic models of inflation due to predicting an overly scale-dependent spectrum, or overproduction of tensor fluctuations \cite{Ade:2015lrj,Ade:2015tva}. Analysis of such CMB experiments typically parametrise the primordial power spectrum $\mathcal{P}_\mathcal{R}$ as a scale-invariant amplitude term $A_s$ plus a power series encoding the scale-dependent deviation from this, such that


\beq
\ln\rpar{\mathcal{P}_\mathcal{R}} = \ln \rpar{A_s} + (n_s - 1) \ln \rpar{\frac{k}{k^*}} + \frac{1}{2} \alpha_s \ln^2 \rpar{\frac{k}{k^*}} + \frac{1}{6} \beta_s \ln^3 \rpar{\frac{k}{k^*}} \ + \ \ldots \, , \label{eq:PSParam}
\eeq

where $n_s$ is called the spectral index, $\alpha_s$ is known as the running of the spectral index, while $\beta_s$ is similarly called the running of the running, and $k^*$ is a reference scale, which is usually taken as $0.05 \ \si{ \mega \parsec^{-1}}$. Further terms can of course be added here, parametrising yet-higher order deviations from scale invariance, but in the most typical analyses of CMB data, $\beta_s$ and any higher order runnings are assumed to be zero. Constraining the parameters with the Planck experiment data, in this case, yields:

\begin{align}
n_s & = 0.9655 \pm 0.0062 \, , \label{eq:standardfit1} \\ 
\alpha_s & = -0.0084 \pm 0.0082 \, , \\
\beta_s & \equiv 0  \, . \label{eq:standardfit3}
\end{align}

This looks largely consistent with the idea that one can safely neglect the higher-order runnings such as $\beta_s$, as it appears likely that even the first running $\alpha_s$ is order of magnitude or two smaller than $(n_s - 1)$. The runnings hence seem to follow a standard hierarchy where each higher order correction is suppressed by a smaller coefficient than the previous term in the expansion. 

To confirm that this is reliable and that we can be satisfied with our parametrisation of $\mathcal{P}_\mathcal{R}$ as a truncated power series, however, one should also consider an analysis which does allow $\beta_s$ to take a nonzero value. One would expect to see that the best fit values given in eqs. (\ref{eq:standardfit1} -- \ref{eq:standardfit3}) are only mildly perturbed by the addition of this extra parameter, and that the best fit value of $\beta_s$ would be largely negligible. This is not the case, though, as one instead finds the best fit for the primordial power spectrum cut off at the $\beta_s$ term is

\begin{align}
n_s & = 0.9569 \pm 0.0077 \, , \\ 
\alpha_s & = 0.011^{+0.014}_{-0.013} \, , \\
\beta_s & = 0.029^{+0.015}_{-0.016} \, .
\end{align}

Here, $\alpha_s$ is still largely consistent with zero, albeit with a slight hint of now being positive and possibly more comparable in size to $(n_s - 1)$. More strikingly, however, is the suggestion that $\beta_s$ is positive at nearly $2 \sigma$, and possibly larger than $\alpha_s$. As well as the simple curiosity that the analysis may be unstable under the inclusion of an additional parameter, the Planck results note that ``Allowing for running of the running provides a better fit to the temperature spectrum at low multipoles''. An extended analysis of the Planck experiment's data detailed in \cite{Cabass:2016ldu}, in which the lensing amplitude $A_L$, density parameter of curvature $\Omega_k$ and neutrino masses $\sum m_\nu$ are independently accounted for, find best fit values of

\begin{align}
n_s & = 0.9582^{+0.0055}_{-0.0054}\, , \\ 
\alpha_s & = 0.011 \pm 0.010\, , \\
\beta_s & = 0.027 \pm 0.013 \, ,
\end{align}

which provides a slightly stronger hint at deviation from the standard hierarchy, with $\beta_s$ now positive at more than $2 \sigma$ and larger than $\alpha_s$ at a greater significance than the Planck analysis. This analysis also finds that the tension in the original Planck analysis with the expectation of $A_L = 1$ and $\Omega_k = 0$ is slightly alleviated. Along with the better fit of the data at low multipoles, these observations suggest that we should take these analyses with $\beta_s \neq 0$ seriously.

The large error bars on these quantities mean that further data and analyses are needed before one can conclusively say that nature is not consistent with an approximately zero $\beta_s$, but, as we will discuss in this article, this also may be a hint of the workings of some interesting physics. While we hence dedicate our attention for the moment to an exploratory study of the theoretical feasibility of a non-standard hierarchy in which $|\beta_s| \gtrapprox |\alpha_s|$, it is also important to contemplate the future prospects for improving constraints on the runnings, and thus becoming able to vindicate or exclude inflationary models based on their hierarchical structure of runnings.
 
Measurements of spectral distortions in the CMB could improve our constraints on the hierarchy of runnings, and could be achieved at a useful significance by the proposed PIXIE experiment \cite{Cabass:2016giw,Chluba:2016bvg,Chluba:2015bqa,Kogut2011}. Similarly, future CMB survey missions such as CORE \cite{DiValentino:2016foa,Finelli:2016cyd} or PRISM \cite{Andre:2013nfa}, 21 cm mapping \cite{Battye:2004re} with an instrument like the Square Kilometre Array \cite{Maartens:2015mra}, or a spectroscopic galaxy survey with the Euclid satellite \cite{Amendola:2016saw} could constrain $\alpha_s$ and $\beta_s$. Particularly with a combination of some or all of these future experiments (or comparable alternatives), as forecasted in \cite{Pourtsidou:2016ctq}, the uncertainty in the runnings could be brought down to $O(10^{-3})$.
 
\section{Predictions of single field slow-roll inflation}

A first question to ask is whether the simplest and most ubiquitous inflationary scenario, in which one canonical scalar field slow-rolls down its potential, can invoke a non-standard hierarchy of runnings in which $|\beta_s| \gtrapprox |\alpha_s|$. The well-known result for the primordial power spectrum in such scenarios is

\beq
\mathcal{P}_\mathcal{R} = \frac{H^2}{8 \pi^2 \epsilon_0} \, , \label{eq:SFSRPS}
\eeq

where $H$ is the Hubble parameter, and $\epsilon_0$ is the first slow-roll parameter with the usual definition

\beq
\epsilon_0 = -\frac{\dot{H}}{H^2} \, ,
\eeq

which has the property that when $\epsilon > 1$, the expansion of the universe ceases to be inflationary. As this is a single field result, all quantities are evaluated at the moment of horizon crossing (typically 50 or 60 $e$-folds before inflation ends, such that $\epsilon_0 \ll 1$) and no significant evolution of the curvature perturbation occurs on superhorizon scales such that the power spectrum at this point is the final power spectrum to a very good approximation. From eq. (\ref{eq:SFSRPS}) we can determine the spectral index and runnings as defined in eq. (\ref{eq:PSParam}), via

\begin{align}
n_s - 1 & = \frac{\bd \ln \mathcal{P}_\mathcal{R}}{ \bd \ln k} = - 2 \epsilon_0 - \epsilon_1 \, , \label{eq:srns} \\
\alpha_s & = \frac{\bd n_s}{ \bd \ln k} = - 2 \epsilon_0 \epsilon_1 - \epsilon_1 \epsilon_2 \, , \\
\beta_s & = \frac{\bd \alpha_s}{ \bd \ln k} = - 2 \epsilon_0 \epsilon_1 (\epsilon_1 + \epsilon_2) - \epsilon_1 \epsilon_2 (\epsilon_2 + \epsilon_3) \, , \label{eq:srbetas}
\end{align}

where we have defined a series of slow-roll parameters $\epsilon_n$ recursively such that

\beq
\epsilon_{n+1} = \frac{\dot{\epsilon}_n}{H \epsilon_n} \, .
\eeq

Once again, at the point of horizon crossing, all of the $\epsilon_n$ must be very small for inflation to be sustained for a sufficient number of $e$-folds. This statement could potentially be circumvented if some of the slow-roll parameters are temporarily enlarged due to, say, a localised feature in the potential \cite{Adams:2001vc,Ashoorioon:2014yua}, but this would need to be fine tuned to occur as observable modes leave the horizon. As a result, while $n_s - 1$ is $O(\epsilon_n) \ll 1$, the runnings $\alpha_s$ and $\beta_s$ are found to be $O(\epsilon_n^2)$ and $O(\epsilon_n^3)$ respectively, and hence progressively smaller. Indeed, given that to fit the data $n_s - 1 \approx -4 \times 10^{-2}$, one would roughly expect $\alpha \approx 10^{-3}$ and $\beta_s \approx 10^{-5}$. This is of course consistent with the analysis in which one assumes $\beta_s \equiv 0$, but harder to reconcile with the extended analyses including a nonzero $\beta_s$ discussed in section \ref{sec:intro}. This is both due to the hierarchy of the magnitude of the runnings and the signs; eqs. (\ref{eq:srns} -- \ref{eq:srbetas}) predict negative runnings (as $\epsilon_n \geq 0$), while the experimental data for $\beta_s \neq 0$ favour positive values of $\alpha_s$ and $\beta_s$ at a significance of up to $2 \sigma$.

The extremely small values of runnings predicted by single field slow-rolling models mean that if the realisation of inflation in nature is of this kind, a conclusive detection of such small running values will be experimentally challenging \cite{Munoz:2016owz}, albeit eventually possible. Correspondingly, the detection of a larger running of the running of, say, $O(10^{-2})$ could be feasible much sooner, with less sensitive experiments, and it is possible that the results quoted in section \ref{sec:intro} are the first signs we are seeing of this. Given this state of affairs, we argue that it is important to begin considering what kind of inflationary scenarios may lead to such unexpectedly large values of the running of the running, such that we are in a position to more readily confront observation with theory should such a detection occur. Relatively little work has been done on understanding $\alpha_s$ \cite{Garcia-Bellido:2014gna,Gariazzo:2016blm,Kohri:2014jma,Peloso:2014oza}, never mind $\beta_s$, which is usually neglected. Nevertheless, with further work, this could potentially prove a powerful constrain on inflationary models. The next section is thus dedicated to an initial exploration of achieving a non-standard hierarchy of runnings theoretically.

\section{Multi-field scenarios and isocurvature perturbations}

Part of the issue preventing the single field models of the previous section from generating large runnings is that the final power spectrum is determined at the moment of horizon crossing where the slow-roll approximation is still strongly valid. Models containing more than one field, in which the superhorizon growth of curvature perturbations due to interaction with isocurvature modes breaks this limitation of single field inflation, are a promising arena in which to investigate the possibility of non-standard runnings of the spectral index. The final power spectrum when this is taken into account is given by \cite{Wands:2002bn,Ashoorioon:2008qr,Lalak:2007vi},

\beq
\mathcal{P}_\mathcal{R} = \mathcal{P}_\mathcal{R}^* \rpar{1 + \TRS^2} \, , \label{eq:trsspec}
\eeq

where $\mathcal{P}_\mathcal{R}^*$ is the horizon-crossing power spectrum (starred quantities will henceforth denote values at horizon crossing) and $\TRS$ is the isocurvature-to-curvature transfer function, which is encodes the superhorizon evolution of perturbations via,

\beq
\TRS = \int_{t^*}^{t_\text{end}} A(t) H(t) \TSS \bd t \, , \label{eq:TRSdef}
\eeq

where the isocurvature-to-isocurvature transfer function is given by,

\beq
\TSS = \exp\rpar{\int_{t^*}^{t_\text{end}} B(t) H(t) \bd t} \, , \label{eq:TSSdef}
\eeq

and $A$ and $B$ are model-dependent functions of background quantities, derived from the perturbed equations of motion of the two-field system \cite{DiMarco:2002eb,DiMarco:2005nq,vandeBruck:2014ata}, and appearing in the large-scale equations of motion as,

\beq
\dot{\mathcal{R}} = A H \mathcal{S} \, , \quad \dot{\mathcal{S}} = B H \mathcal{S} \, .
\eeq

Using the definitions of the spectral index and runnings (the first equalities in eqs. (\ref{eq:srns} -- \ref{eq:srbetas})), we can calculate the corrections due to isocurvature transfer from eq. (\ref{eq:trsspec}) as

\begin{align}
\rpar{n_s - 1} & = \frac{\bd \ln \mathcal{P}_\mathcal{R}}{ \bd \ln k} = \rpar{n_s^* - 1} + \frac{1}{H^*} \frac{\bd \ln \rpar{1+\TRS^2}}{\bd t^*}  \, , \\
\alpha & = \frac{\bd n_s}{\bd \ln k} = \alpha^* + \frac{1}{(H^*)^2} \frac{\bd^2 \ln \rpar{1+\TRS^2}}{\bd (t^*)^2} \, , \\
\beta & = \frac{\bd \alpha}{\bd \ln k} = \beta^* +  \frac{1}{(H^*)^3} \frac{\bd^3 \ln \rpar{1+\TRS^2}}{\bd (t^*)^3} \, ,
\end{align}

where the first term is the spectral index/running/running of the running at horizon crossing and the second term in each equation is the effect of $\TRS$ on the final values of these spectral parameters. In these expressions we have converted derivatives with respect to $k$ into time derivatives at horizon crossing $t^*$ by noting $k = a^* H^*$ such that $\dot{k} = a^* (H^*)^2 (1 - \epsilon_0^*) \approx a^* (H^*)^2$ . The correction terms can be re-expressed by explicitly computing the $t^*$ derivatives of eqs. (\ref{eq:TRSdef} -- \ref{eq:TSSdef}), with which we obtain \cite{vandeBruck:2016rfv}

\begin{align}
n_s - 1 =(n_s-1)^*  - 2 \sin \Theta & \rpar{A^* \cos \Theta + B^* \sin \Theta}\, ,\label{eq:ns} \\
\alpha_s =\alpha_s^*+ 2 \cos \Theta & \rpar{A^* \cos \Theta + B^* \sin \Theta}\rpar{A^* \cos 2 \Theta + B^* \sin 2 \Theta} \, ,\label{eq:alpha}\\
\beta_s =\beta_s^*   -  2 \cos \Theta & \rpar{A^* \cos \Theta + B^* \sin \Theta} \rpar{B^* \cos 2 \Theta - A^* \sin 2 \Theta} \nonumber \\
\times &  \rpar{A^* + 2A^* \cos 2\Theta + 2 B^* \sin 2 \Theta}\, ,\label{eq:beta}
\end{align}

where $\tan \Theta = \TRS$ is the transfer angle. Hence, the spectral index/running/running of the running receive corrections proportional to the model dependent parameters $A^*$ and $B^*$ When $\Theta = 0$, corresponding to the power spectrum remaining the same amplitude despite the interaction with isocurvature modes, $n_s = n_s^*$ due to the factor of $\sin \Theta$, while $\alpha_s$ and $\beta_s$ receive corrections. Similarly, when $\Theta$ approaches $\pi/2$, corresponding to the final power spectrum amplitude being primarily due to isocurvature transfer ($\TRS \gg 1$), $\alpha_s$ and $\beta_s$ are uncorrected while $n_s$ is disturbed from its horizon crossing value. This latter case is undesirable, as it does not help us obtain unconventional runnings, but may ruin the smallness of the spectral index. By contrast, the aforementioned limit of a zero transfer angle is appealing as it does not perturb the spectral index, while allowing potentially large corrections to $\alpha_s$ and $\beta_s$ for appropriate values of $A^*$ and $B^*$. In particular, taking the leading contribution around the point $\Theta \rightarrow 0$, we find,

\begin{align}
n_s - 1 =(n_s-1)^* \, , \\
\alpha_s =\alpha_s^*+ 2 (A^*)^2  \, ,\\
\beta_s =\beta_s^*   -  6 (A^*)^2 B^* \, .
\end{align}

Here, if $B^*$ is sufficiently large and negative, one could amplify $\beta_s$ significantly above its horizon-crossing value. For intermediate values of $\Theta$, one could find cases in which  all three parameters are uncorrected (e.g. $ \rpar{A^* \cos \Theta + B^* \sin \Theta} = 0$) and many cases where all three parameters are altered in different ways due to the different trigonometric dependence on $\Theta$ of each parameter. This shows that the introduction of additional fields, and their resulting isocurvature perturbations, vastly expand the possible phenomenology of the spectral runnings, as long as $A^*$ and $B^*$ are non-negligible.

While it is hence possible for multiple field models to produce non-standard hierarchies of runnings, the simplest model, two canonical, non-interacting fields with minimal potentials (only mass terms) are incapable of achieving this in practice, as for any mass ratio the values of $A^*$ and $B^*$ are small, and thus $\beta_s \approx \beta_s^*$ and so on for any transfer angle. On the other hand, a minimal extension of such a model, described by the action \cite{DiMarco:2002eb,DiMarco:2005nq,Kaiser:2013sna,Schutz:2013fua}

\beq
S = \int \bd^4 x \sqrt{-g} \spar{\frac{1}{2} R - \frac{1}{2} (\partial \phi)^2 + - \frac{1}{2} e^{2 b(\phi)} (\partial \chi)^2 - V(\phi,\chi)} \, ,
\eeq

where the potential does not necessarily need to, but may, include interaction terms such as $g^2 \phi^2 \chi^2$, can, for specific parameters (at least in the $b(\phi) \propto \phi$ case, but likely for a range of non-canonical kinetic terms), achieve non-standard hierarchies that approximately fit the data. \cite{vandeBruck:2016rfv}. It would be interesting for future work to study a range of models based on e.g. supergravity, where a non-minimal K{\"a}hler potential naturally gives rise to such non-canonical kinetic terms, or other physically motivated actions of this form. It is important to note that for some theories of this class, a leading-order slow-roll analysis is not sufficient and it has been found that second order corrections play an important role \cite{vandeBruck:2014ata}.

The presence of isocurvature modes also modifies the standard inflationary consistency relation between the tensor-to-scalar ratio $r$ and the tensor spectral index $n_t$, such that now

\beq
r = - 8 n_t \cos^2 \Theta \, .
\eeq

Future constraints on $r$ and $n_t$ could hence infer the possible values of $\Theta$ in a multi-field scenario, and act as a consistency check with the possible values of $\Theta$ implied by measurements of the runnings in a model, through eqs. (\ref{eq:ns} -- \ref{eq:beta}).

\section{Other Approaches}

As a complementary approach to the above strategy of building more complicated models of inflation to produce a non-standard running hierarchy, one could also consider alternatives to the inflationary paradigm such as Variable Speed of Light (VSL) \cite{Moffat:1992ud} models or model-independent corrections to the inflationary spectra coming from theories of quantum gravity. We will, in this section, briefly cover an example of each of these from the literature to explore the feasibility of achieving non-standard runnings in each case.

\subsection{Variable Speed of Light Cosmology}

VSL models are a popular alternative to inflation which may be physically realised for example in disformally related bimetric theories, which allow the ratio of the speed of light to the speed of gravity to vary \cite{Afshordi:2016guo,Moffat:2004qs,Magueijo:2003gj}. Proponents argue that it is more predictive than inflation, and does not suffer from problems of fine-tuning present in inflation \cite{Moffat:2002nm}. While here we are interested in the early universe effects of a variable speed of light, constraints on and evidence for VSL cosmologies at late times \cite{vandeBruck:2016cnh,Salzano:2016pny} have also been studied. In the bimetric VSL model of Moffat \cite{Moffat:2004qs,Moffat:2013nsa}, the power spectrum takes the form

\beq
\mathcal{P}_\mathcal{R} \propto  \ln^2 \rpar{A k^3} \, ,
\eeq

where $A$ is a constant depending on model parameters, fixed by normalising the amplitude of the power spectrum. From this one obtains,

\begin{align}
n_s = 1 + \frac{6}{\ln \rpar{A k^3}} & \approx 0.96 \, , \\
\alpha_s = -\frac{1}{2} \rpar{n_s-1}^2 & \approx -8 \times 10^{-4}  \, ,\\
\beta_s = \frac{1}{2} \rpar{n_s -1}^3 & \approx -3 \times 10^{-5} \, .
\end{align}

This is largely consistent with the Planck analysis assuming $\beta_s \equiv 0$ - a standard hierarchy of runnings. The relations  $\alpha_s \propto (n_s -1)^2$ and $\beta_s \propto (n_s-1)^3$ are characteristic of standard hierarchies and very similar to the predictions of single field slow roll inflation (as $n_s-1 \propto \epsilon$, $\alpha_s \propto \epsilon^2$ and so on). Similarly to the expectation of fiducial models of inflation (\ref{eq:srbetas}), this predicts $\beta_s < 0$. Overall, this model is largely indistinguishable from the most simplistic versions of inflation and as a result, to achieve consistency with a more certain detection of an alternative hierarchy with $\beta_s \approx O(10^{-2})$ this basic realisation of VSL would have to be extended, much like the most basic inflationary scenarios.

\subsection{Cosmological Perturbations in Quantum Gravity}

Theories of quantum gravity predict modifications of the dynamics of cosmological perturbations and hence corrections to the inflationary power spectra. In canonical quantum gravity, for example, it has been shown \cite{Brizuela:2015tzl,Brizuela:2016gnz} that the corrected power spectrum takes the form

\beq
\mathcal{P}_\mathcal{R} = \mathcal{P}_\mathcal{R}^{(0)} (1 + \Delta) \, , \label{eq:cqgcorrection}
\eeq

where $\mathcal{P}_\mathcal{R}^{(0)}$ is the usual, uncorrected spectrum and

\beq
\Delta = 0.988 \frac{H^2}{M_\text{pl}^2} \rpar{\frac{\bar{k}}{k}}^3  + O(\epsilon) \, ,
\eeq

is the leading order (in slow-roll) correction due to quantum gravity effects, in which $\bar{k}$ is an introduced length scale of the theory. It is noted that this kind of correction term does not lend itself well to the common parametrisation of the spectrum in eq. (\ref{eq:PSParam}). In particular, the explicit dependence on $k$ will lead to terms in the spectral index and runnings which are not suppressed by increasing orders of slow-roll parameters, and hence could be of similar order. This is interesting for our purposes, as it could quite naturally explain a deviation from the expected hierarchy of runnings, without having to invoke complicated models of inflation. On a similar note, theories of loop quantum gravity such as \cite{Bojowald:2011iq} have been shown to produce spectra with higher order runnings all of a similar magnitude. In general, the field of quantum gravity would hence seem to be an interesting area to further study in the context of going beyond the usual slow-roll hierarchy of runnings.

Explicitly calculating the spectral index and its runnings from eq. (\ref{eq:cqgcorrection}), we obtain,

\begin{align}
n_s - 1 = (n_s - 1)^{(0)} - \frac{3 \Delta}{\rpar{1 + \Delta}} + O(\epsilon) \, , \\
\alpha_s = \alpha_s^{(0)} + \frac{9 \Delta}{\rpar{1 + \Delta}^2} + O(\epsilon) \, ,\\
\beta_s = \beta_s^{(0)} + \frac{27 \Delta \rpar{\Delta - 1}}{\rpar{1 + \Delta}^3} + O(\epsilon)  \, . \label{eq:cqgbeta}
\end{align}

In \cite{Brizuela:2016gnz} the authors explicitly compute the spectral index and running for the case where $H \approx 10^{-5} M_\text{Pl}$ for a fiducial scale $\bar{k}$ to find that in such models, when $\Delta$ is $O(10^{-10})$, the corrections to the spectral parameters are likely unobservably small. The correction to $\alpha$ is however greater in magnitude than the one to $n_s$, confirming the expectation that this quantum gravity correction would not fit the standard hierarchy of runnings. Indeed, in the limit of small $\Delta$, the leading order corrections are

\begin{align}
n_s - 1 \approx (n_s - 1)^{(0)} - 3 \Delta \, , \\
\alpha_s = \alpha_s^{(0)} + 9 \Delta \, ,\\
\beta_s = \beta_s^{(0)} - 27 \Delta  \, ,
\end{align}

such that each subsequent parameter is a factor of 3 greater in magnitude than the previous one. One could also consider alternative scenarios in which $\Delta$ is not small (e.g. when $H/M_\text{Pl}$ is larger for some reason, or for a larger choice of $\bar{k}$) and the final power spectrum (\ref{eq:cqgcorrection}) receives a greater contribution from quantum gravity effects. In such a scenario, to obtain a positive correction to $\beta_s$ as is found in the analysis of \cite{Cabass:2016ldu}, from (\ref{eq:cqgbeta}), one needs $\Delta > 1$ (though we note in this regime higher order corrections would also likely be important). For such a large $\Delta$,  $n_s - 1$ and $\alpha_s$ are corrected by $O(1)$, which likely ruins their fit to the data unless the uncorrected values ($(n_s - 1)^{(0)}$ , etc.) are abnormal in the first place. 

To conclude, while quantum gravity effects may be capable of naturally breaking the hierarchy of runnings, this particular realisation appears unsuitable for achieving the specific modified hierarchy that recent analyses have hinted at, as the corrections are either too small (when $\Delta \ll 1$), of the wrong sign for $\beta_s$ (when $0 < \Delta < 1$), or too large in $n_s$ and $\alpha_s$ (when $\Delta > 1$).

\section{Conclusion}

Given recent observational hints of an unexpectedly large running of the running ($\beta_s$), and the resulting non-standard hierarchy of runnings, we undertook a theoretical exploration of the possible implications of this and what kinds of physical models might be able to realise such a hint if it is later confirmed by further analyses and experiments. We find that single field slow-roll inflation would be very difficult to reconcile with such a hierarchy. Models of multi-field inflation with non-canonical kinetic terms, due to the influence of isocurvature perturbations on the spectrum, are however capable of generating this kind of hierarchy. In any case, future measurements of the running and running of the running may provide a powerful discriminator between inflationary models. Variable speed of light models do not alleviate the problem, with the simplest realisations predicting the same kind of running hierarchy as single field slow-roll inflation. Quantum gravity corrections to the primordial power spectra, however, seem to have the right kind of qualitative properties to explain an unusually large running of the running, taking a form which does not lend itself well to comparison with the usual parametrisation. Future work in this direction may hence be able to shed some light on the nature of quantum gravity effects.

%

\vspace{6pt} 


\acknowledgments{I would like to thank Carsten van de Bruck for the many invaluable discussions we had and continue to have on this subject matter. I acknowledge receipt of an STFC studentship throughout the creation of this work.}


\conflictofinterests{The authors declare no conflict of interest.} 

%

%

\bibliographystyle{mdpi}

\bibliography{runningproceedingsrefs}

\begin{thebibliography}{-------}
\providecommand{\natexlab}[1]{#1}

\bibitem[Guth(1981)]{Guth:1980zm}
Guth, A.H.
\newblock {The Inflationary Universe: A Possible Solution to the Horizon and
  Flatness Problems}.
\newblock {\em Phys. Rev.} {\bf 1981}, {\em D23},~347--356.

\bibitem[Linde(1982)]{Linde:1981mu}
Linde, A.D.
\newblock {A New Inflationary Universe Scenario: A Possible Solution of the
  Horizon, Flatness, Homogeneity, Isotropy and Primordial Monopole Problems}.
\newblock {\em Phys. Lett.} {\bf 1982}, {\em B108},~389--393.

\bibitem[Escudero \em{et~al.}(2016)Escudero, Ram\'{i}rez, Boubekeur, Giusarma,
  and Mena]{Escudero:2015wba}
Escudero, M.; Ram\'{i}rez, H.; Boubekeur, L.; Giusarma, E.; Mena, O.
\newblock {The present and future of the most favoured inflationary models
  after $Planck$ 2015}.
\newblock {\em JCAP} {\bf 2016}, {\em 1602},~020,
  \href{http://xxx.lanl.gov/abs/1509.05419}{{\normalfont
  [arXiv:astro-ph.CO/1509.05419]}}.

\bibitem[Martin \em{et~al.}(2014)Martin, Ringeval, and Vennin]{Martin:2013tda}
Martin, J.; Ringeval, C.; Vennin, V.
\newblock {Encyclopaedia Inflationaris}.
\newblock {\em Phys. Dark Univ.} {\bf 2014}, {\em 5-6},~75--235,
  \href{http://xxx.lanl.gov/abs/1303.3787}{{\normalfont
  [arXiv:astro-ph.CO/1303.3787]}}.

\bibitem[Ade \em{et~al.}(2015{\natexlab{a}})Ade et~al.]{Ade:2015lrj}
Ade, P.A.R.; others.
\newblock {Planck 2015 results. XX. Constraints on inflation} {\bf 2015}.
\newblock  \href{http://xxx.lanl.gov/abs/1502.02114}{{\normalfont
  [arXiv:astro-ph.CO/1502.02114]}}.

\bibitem[Ade \em{et~al.}(2015{\natexlab{b}})Ade et~al.]{Ade:2015tva}
Ade, P.A.R.; others.
\newblock {Joint Analysis of BICEP2/$Keck Array$ and $Planck$ Data}.
\newblock {\em Phys. Rev. Lett.} {\bf 2015}, {\em 114},~101301,
  \href{http://xxx.lanl.gov/abs/1502.00612}{{\normalfont
  [arXiv:astro-ph.CO/1502.00612]}}.

\bibitem[Cabass \em{et~al.}(2016{\natexlab{a}})Cabass, Di~Valentino,
  Melchiorri, Pajer, and Silk]{Cabass:2016ldu}
Cabass, G.; Di~Valentino, E.; Melchiorri, A.; Pajer, E.; Silk, J.
\newblock {Running the running} {\bf 2016}.
\newblock  \href{http://xxx.lanl.gov/abs/1605.00209}{{\normalfont
  [arXiv:astro-ph.CO/1605.00209]}}.

\bibitem[Cabass \em{et~al.}(2016{\natexlab{b}})Cabass, Melchiorri, and
  Pajer]{Cabass:2016giw}
Cabass, G.; Melchiorri, A.; Pajer, E.
\newblock {$\mu$ distortions or running: A guaranteed discovery from CMB
  spectrometry}.
\newblock {\em Phys. Rev.} {\bf 2016}, {\em D93},~083515,
  \href{http://xxx.lanl.gov/abs/1602.05578}{{\normalfont
  [arXiv:astro-ph.CO/1602.05578]}}.

\bibitem[Chluba(2016)]{Chluba:2016bvg}
Chluba, J.
\newblock {Which spectral distortions does $\Lambda$CDM actually predict?} {\bf
  2016}.
\newblock  \href{http://xxx.lanl.gov/abs/1603.02496}{{\normalfont
  [arXiv:astro-ph.CO/1603.02496]}}.

\bibitem[Chluba \em{et~al.}(2015)Chluba, Hamann, and Patil]{Chluba:2015bqa}
Chluba, J.; Hamann, J.; Patil, S.P.
\newblock {Features and New Physical Scales in Primordial Observables: Theory
  and Observation}.
\newblock {\em Int. J. Mod. Phys.} {\bf 2015}, {\em D24},~1530023,
  \href{http://xxx.lanl.gov/abs/1505.01834}{{\normalfont
  [arXiv:astro-ph.CO/1505.01834]}}.

\bibitem[{Kogut} \em{et~al.}(2011){Kogut}, {Fixsen}, {Chuss}, {Dotson}, {Dwek},
  {Halpern}, {Hinshaw}, {Meyer}, {Moseley}, {Seiffert}, {Spergel}, and
  {Wollack}]{Kogut2011}
{Kogut}, A.; {Fixsen}, D.J.; {Chuss}, D.T.; {Dotson}, J.; {Dwek}, E.;
  {Halpern}, M.; {Hinshaw}, G.F.; {Meyer}, S.M.; {Moseley}, S.H.; {Seiffert},
  M.D.; {Spergel}, D.N.; {Wollack}, E.J.
\newblock {The Primordial Inflation Explorer (PIXIE): a nulling polarimeter for
  cosmic microwave background observations}.
\newblock {\em JCAP} {\bf 2011}, {\em 7},~025,
  \href{http://xxx.lanl.gov/abs/1105.2044}{{\normalfont [1105.2044]}}.

\bibitem[Di~Valentino \em{et~al.}(2016)Di~Valentino
  et~al.]{DiValentino:2016foa}
Di~Valentino, E.; others.
\newblock {Exploring Cosmic Origins with CORE: Cosmological Parameters} {\bf
  2016}.
\newblock  \href{http://xxx.lanl.gov/abs/1612.00021}{{\normalfont
  [arXiv:astro-ph.CO/1612.00021]}}.

\bibitem[Finelli \em{et~al.}(2016)Finelli et~al.]{Finelli:2016cyd}
Finelli, F.; others.
\newblock {Exploring Cosmic Origins with CORE: Inflation} {\bf 2016}.
\newblock  \href{http://xxx.lanl.gov/abs/1612.08270}{{\normalfont
  [arXiv:astro-ph.CO/1612.08270]}}.

\bibitem[Andr\'{e} \em{et~al.}(2014)Andr\'{e} et~al.]{Andre:2013nfa}
Andr\'{e}, P.; others.
\newblock {PRISM (Polarized Radiation Imaging and Spectroscopy Mission): An
  Extended White Paper}.
\newblock {\em JCAP} {\bf 2014}, {\em 1402},~006,
  \href{http://xxx.lanl.gov/abs/1310.1554}{{\normalfont
  [arXiv:astro-ph.CO/1310.1554]}}.

\bibitem[Battye \em{et~al.}(2004)Battye, Davies, and Weller]{Battye:2004re}
Battye, R.A.; Davies, R.D.; Weller, J.
\newblock {Neutral hydrogen surveys for high redshift galaxy clusters and
  proto-clusters}.
\newblock {\em Mon. Not. Roy. Astron. Soc.} {\bf 2004}, {\em 355},~1339--1347,
  \href{http://xxx.lanl.gov/abs/astro-ph/0401340}{{\normalfont
  [arXiv:astro-ph/astro-ph/0401340]}}.

\bibitem[Maartens \em{et~al.}(2015)Maartens, Abdalla, Jarvis, and
  Santos]{Maartens:2015mra}
Maartens, R.; Abdalla, F.B.; Jarvis, M.; Santos, M.G.
\newblock {Overview of Cosmology with the SKA}.
\newblock {\em PoS} {\bf 2015}, {\em AASKA14},~016,
  \href{http://xxx.lanl.gov/abs/1501.04076}{{\normalfont
  [arXiv:astro-ph.CO/1501.04076]}}.

\bibitem[Amendola \em{et~al.}(2016)Amendola et~al.]{Amendola:2016saw}
Amendola, L.; others.
\newblock {Cosmology and Fundamental Physics with the Euclid Satellite} {\bf
  2016}.
\newblock  \href{http://xxx.lanl.gov/abs/1606.00180}{{\normalfont
  [arXiv:astro-ph.CO/1606.00180]}}.

\bibitem[Pourtsidou(2016)]{Pourtsidou:2016ctq}
Pourtsidou, A.
\newblock {Synergistic tests of inflation} {\bf 2016}.
\newblock  \href{http://xxx.lanl.gov/abs/1612.05138}{{\normalfont
  [arXiv:astro-ph.CO/1612.05138]}}.

\bibitem[Adams \em{et~al.}(2001)Adams, Cresswell, and Easther]{Adams:2001vc}
Adams, J.A.; Cresswell, B.; Easther, R.
\newblock {Inflationary perturbations from a potential with a step}.
\newblock {\em Phys. Rev.} {\bf 2001}, {\em D64},~123514,
  \href{http://xxx.lanl.gov/abs/astro-ph/0102236}{{\normalfont
  [arXiv:astro-ph/astro-ph/0102236]}}.

\bibitem[Ashoorioon \em{et~al.}(2014)Ashoorioon, van~de Bruck, Millington, and
  Vu]{Ashoorioon:2014yua}
Ashoorioon, A.; van~de Bruck, C.; Millington, P.; Vu, S.
\newblock {Effect of transitions in the Planck mass during inflation on
  primordial power spectra}.
\newblock {\em Phys. Rev.} {\bf 2014}, {\em D90},~103515,
  \href{http://xxx.lanl.gov/abs/1406.5466}{{\normalfont
  [arXiv:astro-ph.CO/1406.5466]}}.

\bibitem[Mu\~{n}oz \em{et~al.}(2016)Mu\~{n}oz, Kovetz, Raccanelli,
  Kamionkowski, and Silk]{Munoz:2016owz}
Mu\~{n}oz, J.B.; Kovetz, E.D.; Raccanelli, A.; Kamionkowski, M.; Silk, J.
\newblock {Towards a measurement of the spectral runnings} {\bf 2016}.
\newblock  \href{http://xxx.lanl.gov/abs/1611.05883}{{\normalfont
  [arXiv:astro-ph.CO/1611.05883]}}.

\bibitem[Garcia-Bellido and Roest(2014)]{Garcia-Bellido:2014gna}
Garcia-Bellido, J.; Roest, D.
\newblock {Large-$N$ running of the spectral index of inflation}.
\newblock {\em Phys. Rev.} {\bf 2014}, {\em D89},~103527,
  \href{http://xxx.lanl.gov/abs/1402.2059}{{\normalfont
  [arXiv:astro-ph.CO/1402.2059]}}.

\bibitem[Gariazzo \em{et~al.}(2016)Gariazzo, Mena, Ramirez, and
  Boubekeur]{Gariazzo:2016blm}
Gariazzo, S.; Mena, O.; Ramirez, H.; Boubekeur, L.
\newblock {Primordial power spectrum features in phenomenological descriptions
  of inflation} {\bf 2016}.
\newblock  \href{http://xxx.lanl.gov/abs/1606.00842}{{\normalfont
  [arXiv:astro-ph.CO/1606.00842]}}.

\bibitem[Kohri and Matsuda(2015)]{Kohri:2014jma}
Kohri, K.; Matsuda, T.
\newblock {Ambiguity in running spectral index with an extra light field during
  inflation}.
\newblock {\em JCAP} {\bf 2015}, {\em 1502},~019,
  \href{http://xxx.lanl.gov/abs/1405.6769}{{\normalfont
  [arXiv:astro-ph.CO/1405.6769]}}.

\bibitem[Peloso \em{et~al.}(2014)Peloso, Sorbo, and Tasinato]{Peloso:2014oza}
Peloso, M.; Sorbo, L.; Tasinato, G.
\newblock {A falsely fat curvaton with an observable running of the spectral
  tilt}.
\newblock {\em JCAP} {\bf 2014}, {\em 1406},~040,
  \href{http://xxx.lanl.gov/abs/1401.7136}{{\normalfont
  [arXiv:astro-ph.CO/1401.7136]}}.

\bibitem[Wands \em{et~al.}(2002)Wands, Bartolo, Matarrese, and
  Riotto]{Wands:2002bn}
Wands, D.; Bartolo, N.; Matarrese, S.; Riotto, A.
\newblock {An Observational test of two-field inflation}.
\newblock {\em Phys. Rev.} {\bf 2002}, {\em D66},~043520,
  \href{http://xxx.lanl.gov/abs/astro-ph/0205253}{{\normalfont
  [arXiv:astro-ph/astro-ph/0205253]}}.

\bibitem[Ashoorioon \em{et~al.}(2009)Ashoorioon, Krause, and
  Turzynski]{Ashoorioon:2008qr}
Ashoorioon, A.; Krause, A.; Turzynski, K.
\newblock {Energy Transfer in Multi Field Inflation and Cosmological
  Perturbations}.
\newblock {\em JCAP} {\bf 2009}, {\em 0902},~014,
  \href{http://xxx.lanl.gov/abs/0810.4660}{{\normalfont
  [arXiv:hep-th/0810.4660]}}.

\bibitem[Lalak \em{et~al.}(2007)Lalak, Langlois, Pokorski, and
  Turzynski]{Lalak:2007vi}
Lalak, Z.; Langlois, D.; Pokorski, S.; Turzynski, K.
\newblock {Curvature and isocurvature perturbations in two-field inflation}.
\newblock {\em JCAP} {\bf 2007}, {\em 0707},~014,
  \href{http://xxx.lanl.gov/abs/0704.0212}{{\normalfont
  [arXiv:hep-th/0704.0212]}}.

\bibitem[Di~Marco \em{et~al.}(2003)Di~Marco, Finelli, and
  Brandenberger]{DiMarco:2002eb}
Di~Marco, F.; Finelli, F.; Brandenberger, R.
\newblock {Adiabatic and isocurvature perturbations for multifield generalized
  Einstein models}.
\newblock {\em Phys. Rev.} {\bf 2003}, {\em D67},~063512,
  \href{http://xxx.lanl.gov/abs/astro-ph/0211276}{{\normalfont
  [arXiv:astro-ph/astro-ph/0211276]}}.

\bibitem[Di~Marco and Finelli(2005)]{DiMarco:2005nq}
Di~Marco, F.; Finelli, F.
\newblock {Slow-roll inflation for generalized two-field Lagrangians}.
\newblock {\em Phys. Rev.} {\bf 2005}, {\em D71},~123502,
  \href{http://xxx.lanl.gov/abs/astro-ph/0505198}{{\normalfont
  [arXiv:astro-ph/astro-ph/0505198]}}.

\bibitem[van~de Bruck and Robinson(2014)]{vandeBruck:2014ata}
van~de Bruck, C.; Robinson, M.
\newblock {Power Spectra beyond the Slow Roll Approximation in Theories with
  Non-Canonical Kinetic Terms}.
\newblock {\em JCAP} {\bf 2014}, {\em 1408},~024,
  \href{http://xxx.lanl.gov/abs/1404.7806}{{\normalfont
  [arXiv:astro-ph.CO/1404.7806]}}.

\bibitem[van~de Bruck and Longden(2016)]{vandeBruck:2016rfv}
van~de Bruck, C.; Longden, C.
\newblock {Running of the Running and Entropy Perturbations During Inflation}.
\newblock {\em Phys. Rev.} {\bf 2016}, {\em D94},~021301,
  \href{http://xxx.lanl.gov/abs/1606.02176}{{\normalfont
  [arXiv:astro-ph.CO/1606.02176]}}.

\bibitem[Kaiser and Sfakianakis(2014)]{Kaiser:2013sna}
Kaiser, D.I.; Sfakianakis, E.I.
\newblock {Multifield Inflation after Planck: The Case for Nonminimal
  Couplings}.
\newblock {\em Phys. Rev. Lett.} {\bf 2014}, {\em 112},~011302,
  \href{http://xxx.lanl.gov/abs/1304.0363}{{\normalfont
  [arXiv:astro-ph.CO/1304.0363]}}.

\bibitem[Schutz \em{et~al.}(2014)Schutz, Sfakianakis, and
  Kaiser]{Schutz:2013fua}
Schutz, K.; Sfakianakis, E.I.; Kaiser, D.I.
\newblock {Multifield Inflation after Planck: Isocurvature Modes from
  Nonminimal Couplings}.
\newblock {\em Phys. Rev.} {\bf 2014}, {\em D89},~064044,
  \href{http://xxx.lanl.gov/abs/1310.8285}{{\normalfont
  [arXiv:astro-ph.CO/1310.8285]}}.

\bibitem[Moffat(1993)]{Moffat:1992ud}
Moffat, J.W.
\newblock {Superluminary universe: A Possible solution to the initial value
  problem in cosmology}.
\newblock {\em Int. J. Mod. Phys.} {\bf 1993}, {\em D2},~351--366,
  \href{http://xxx.lanl.gov/abs/gr-qc/9211020}{{\normalfont
  [arXiv:gr-qc/gr-qc/9211020]}}.

\bibitem[Afshordi and Magueijo(2016)]{Afshordi:2016guo}
Afshordi, N.; Magueijo, J.
\newblock {The critical geometry of a thermal big bang}.
\newblock {\em Phys. Rev.} {\bf 2016}, {\em D94},~101301,
  \href{http://xxx.lanl.gov/abs/1603.03312}{{\normalfont
  [arXiv:gr-qc/1603.03312]}}.

\bibitem[Moffat(2005)]{Moffat:2004qs}
Moffat, J.W.
\newblock {Variable speed of light cosmology and bimetric gravity: An
  Alternative to standard inflation}.
\newblock {\em Int. J. Mod. Phys.} {\bf 2005}, {\em A20},~1155--1162,
  \href{http://xxx.lanl.gov/abs/gr-qc/0404066}{{\normalfont
  [arXiv:gr-qc/gr-qc/0404066]}}.

\bibitem[Magueijo(2003)]{Magueijo:2003gj}
Magueijo, J.
\newblock {New varying speed of light theories}.
\newblock {\em Rept. Prog. Phys.} {\bf 2003}, {\em 66},~2025,
  \href{http://xxx.lanl.gov/abs/astro-ph/0305457}{{\normalfont
  [arXiv:astro-ph/astro-ph/0305457]}}.

\bibitem[Moffat(2002)]{Moffat:2002nm}
Moffat, J.W.
\newblock {Variable speed of light cosmology: An Alternative to inflation} {\bf
  2002}.
\newblock  \href{http://xxx.lanl.gov/abs/hep-th/0208122}{{\normalfont
  [arXiv:hep-th/hep-th/0208122]}}.

\bibitem[van~de Bruck \em{et~al.}(2016)van~de Bruck, Burrage, and
  Morrice]{vandeBruck:2016cnh}
van~de Bruck, C.; Burrage, C.; Morrice, J.
\newblock {Vacuum Cherenkov radiation and bremsstrahlung from disformal
  couplings}.
\newblock {\em JCAP} {\bf 2016}, {\em 1608},~003,
  \href{http://xxx.lanl.gov/abs/1605.03567}{{\normalfont
  [arXiv:gr-qc/1605.03567]}}.

\bibitem[Salzano and Dabrowski(2016)]{Salzano:2016pny}
Salzano, V.; Dabrowski, M.P.
\newblock {Statistical hierarchy of varying speed of light cosmologies} {\bf
  2016}.
\newblock  \href{http://xxx.lanl.gov/abs/1612.06367}{{\normalfont
  [arXiv:astro-ph.CO/1612.06367]}}.

\bibitem[Moffat(2013)]{Moffat:2013nsa}
Moffat, J.W.
\newblock {Bimetric Gravity, Variable Speed of Light Cosmology and Planck2013}
  {\bf 2013}.
\newblock  \href{http://xxx.lanl.gov/abs/1306.5470}{{\normalfont
  [arXiv:gr-qc/1306.5470]}}.

\bibitem[Brizuela \em{et~al.}(2016{\natexlab{a}})Brizuela, Kiefer, and
  Kraemer]{Brizuela:2015tzl}
Brizuela, D.; Kiefer, C.; Kraemer, M.
\newblock {Quantum-gravitational effects on gauge-invariant scalar and tensor
  perturbations during inflation: The de Sitter case}.
\newblock {\em Phys. Rev.} {\bf 2016}, {\em D93},~104035,
  \href{http://xxx.lanl.gov/abs/1511.05545}{{\normalfont
  [arXiv:gr-qc/1511.05545]}}.

\bibitem[Brizuela \em{et~al.}(2016{\natexlab{b}})Brizuela, Kiefer, and
  Kraemer]{Brizuela:2016gnz}
Brizuela, D.; Kiefer, C.; Kraemer, M.
\newblock {Quantum-gravitational effects on gauge-invariant scalar and tensor
  perturbations during inflation: The slow-roll approximation}.
\newblock {\em Phys. Rev.} {\bf 2016}, {\em D94},~123527,
  \href{http://xxx.lanl.gov/abs/1611.02932}{{\normalfont
  [arXiv:gr-qc/1611.02932]}}.

\bibitem[Bojowald \em{et~al.}(2011)Bojowald, Calcagni, and
  Tsujikawa]{Bojowald:2011iq}
Bojowald, M.; Calcagni, G.; Tsujikawa, S.
\newblock {Observational test of inflation in loop quantum cosmology}.
\newblock {\em JCAP} {\bf 2011}, {\em 1111},~046,
  \href{http://xxx.lanl.gov/abs/1107.1540}{{\normalfont
  [arXiv:gr-qc/1107.1540]}}.

\end{thebibliography}

\end{document}